\documentclass{optica-article}

\journal{oe}

\usepackage{lineno}
\begin{document}

\title{Optimizing the generation of polarization squeezed light in nonlinear optical fibers driven by femtosecond pulses}

\author{A.~V.~Andrianov,\authormark{1} N.~A.~Kalinin,\authormark{1,2} A.~A.~Sorokin, \authormark{1} E.~A.~Anashkina,\authormark{1,3}  L.~L.~S\'{a}nchez-Soto,\authormark{2,4,*}  J.~F.~Corney,\authormark{5} and G.~Leuchs\authormark{1,2,6}}

\address{
\authormark{1}Institute of Applied Physics of the Russian Academy of Sciences, Nizhny Novgorod 603950, Russia\\
\authormark{2}Max Planck Institute for the Science of Light,  91058 Erlangen, Germany\\
\authormark{3}Advanced School of General and Applied Physics, Lobachevsky State University of Nizhny Novgorod, Nizhny Novgorod 603022, Russia\\
\authormark{4}Departamento de \'Optica, Facultad de F\'isica, Universidad Complutense, 28040 Madrid, Spain\\
\authormark{5}School of Mathematics and Physics, University of Queensland, Brisbane, Queensland 4072, Australia \\
\authormark{6}Department of Physics, University of Erlangen-Nuremberg, 91058 Erlangen, Germany}

\email{\authormark{*}lsanchez@fis.ucm.es} %% email address is required; see note below about the corresponding author designation



\begin{abstract}
Bright squeezed light can be generated in optical fibers utilizing the Kerr effect for ultrashort laser pulses. However, pulse propagation in a fiber is subject to nonconservative effects that deteriorate the squeezing. Here, we analyze two-mode polarization squeezing, which is SU(2)-invariant, robust against technical perturbations, and can be generated in a polarization-maintaining fiber. We perform a rigorous numerical optimization of the process and the pulse parameters using our advanced model of quantum pulse evolution in the fiber that includes various nonconservative effects and real fiber data. Numerical results are consistent with experimental results. 
\end{abstract}

%%%%%%%%%%%%%%%%%%%%%%%%%%  body  %%%%%%%%%%%%%%%%%%%%%%%%%%
\section{Introduction}

Squeezed light is one of the most important resources in quantum optics with many existing and foreseen applications, including improving the sensitivity and precision of optical metrology, quantum communications, and quantum computing with continuous variables~\cite{Andersen2016, Lvovsky2014, Pan2020, Lawrie2019}. Squeezed light, as a theoretical concept, has been studied for a long time (see, e.g.~\cite{Dodonov2002} for a review). However, the first experimental observation of squeezing was made in 1986~\cite{Slusher}. Since then, the experimental methods  have improved significantly, and quantum squeezing has already become a useful technology for applications. For example, modern gravitational wave detectors use squeezed light to enhance the sensitivity and increase the observable range in space~\cite{Lough2021}. Squeezed light plays an important role in modern theoretical studies; e.g., quantum cavity electrodynamics, quantum phase transitions~\cite{ZhuPRL2020}, and symmetry breaking in quantum systems~\cite{PerinaSYM2019}. Squeezed light can be generated by using various optical nonlinearities (see, e.g.~\cite{Andersen2016} for a review), including second-order nonlinear processes, such as parametric down-conversion and oscillations~\cite{Wu, Vahlbruch}, parametric up-conversion~\cite{Sizmann, Mehmet}, third-order nonlinearity in atomic vapors and fibers, and also by direct intensity noise reduction by driving semiconductors lasers with extremely low-noise current source~\cite{Machida}. 

In this work, we concentrate on the optical Kerr effect, which can produce squeezing in amorphous media and it is not limited by phase-matching conditions, thus providing larger bandwidth. In the simplest scheme, a coherent state of light is launched into the nonlinear Kerr medium. Amplitude-phase correlations are induced because the nonlinear phase shift is proportional to the intensity. These correlations result in the formation of a squeezed Wigner distribution with elliptic contours in phase space, in contrast to the rotationally symmetric Gaussian distribution of the initial coherent state. The full quantum treatment shows that the Kerr interaction leads to a non-Gaussian periodic dynamics with the appearance of ``cat states" and recurrence to the initial coherent state~\cite{Milburn, Stobinska}. However, for reasonable values of nonlinearities, light power and loss-limited distances, the Gaussian approximation can be used within a large margin.  

In the first fiber experiment continuous-wave (CW) light was used, and less than 1dB squeezing was achieved in 114-m long fiber~\cite{Shelby}. This experiment required enormous efforts to overcome destructive effects of losses and noise induced by Brillouin scattering on the thermally excited guided acoustic phonons in the fiber (GAWBS--guided acoustic waves Brillouin scattering) accumulated over the long fiber. It was then proposed to use pulsed light because it is much easier to achieve high-peak power, keeping the average power at a moderate level, thus requiring much shorter fibers and greatly reducing the effect of losses and GAWBS. Although most of the following fiber squeezing studies rely on short pulses, we note that in modern fibers based on glasses with very high nonlinearity and good transparency, e.g. chalcogenide and tellurite glasses, CW or long-pulse squeezing may worth revisiting~\cite{Anashkina2020, Anashkina2021,Sorokin2022}. A quantum theory of pulse propagation in dispersive nonlinear media~\cite{Carter1987} suggested that quadrature squeezing can be achieved for pulsed light, especially for solitons that preserve their shape and peak intensity over long distances despite dispersion. Early experiments utilized both nonsoliton~\cite{Bergman91,Bergman94} as well as soliton pulse propagation~\cite{Rosenbluh1991, Drummond1993}.

One obstacle in using Kerr squeezing is that the squeezed ellipse is tilted in phase space with respect to the mean vector of the field amplitude so that the output quantum state is not amplitude-squeezed, which hinders direct detection of the reduced noise with power detectors. Several methods to overcome this obstacle were proposed, such as using reflection from a highly dispersive cavity~\cite{Shelby} or employing two-mode squeezing in Sagnac-type and Mach-Zehnder-type fiber interferometers to facilitate heterodyne detection~\cite{Kitagawa1986, Bergman91, Rosenbluh1991, Drummond1993, Schmitt1998, Fiorentino2001}. Symmetric Sagnac interferometers~\cite{Rosenbluh1991, Drummond1993} producing nearly vacuum squeezed state, as well as asymmetric interferometers producing bright coherent squeezed states were used~\cite{Schmitt1998,Krylov98}. Another approach relies on the spectral filtering of the pulse after nonlinear propagation, which converts noise correlations between different spectral bands into directly detectable amplitude squeezing~\cite{Friberg1996}. 
%However, these methods are not easy to set up and maintain under the influence of technical disturbances. 
One of the most robust techniques relies on squeezing of the uncertainty of the polarization state. By generating two squeezed beams in two polarization modes of a polarization-maintaining fiber and appropriately transforming the output polarization state, the reduced uncertainty of the polarization state can be directly measured by power detectors~\cite{Heersink05, Dong2008, Corney2008, Hosaka2015}. The best squeezing achieved so far with fibers was observed in such a system~\cite{Dong2008}. 

Ultrashort pulses propagating in fibers are susceptible to nonconservative effects of spontaneous and stimulated Raman scattering. It was quickly recognized~\cite{Carter91} and tested in experiments and simulations\cite{Corney2008, Dong2008} that the Raman effect is one of the most important factors limiting squeezing in optical fibers for ultrashort pulses. Whereas electronic Kerr nonlinearity is not sensitive to the pulse duration, the delayed Raman contribution is. It is known that the influence of Raman on the classical properties of ultrashort fiber solitons scales with the pulse duration, being much more pronounced for shorter pulses. This suggests that increasing the pulse duration may also help reduce detrimental Raman contribution to quantum squeezing. However, the comprehensive analysis of pulsed Kerr squeezing and optimization over the full set of pulse parameters has not been done yet. In this work we perform rigorous numerical simulations to test the dependence of the squeezing on the pulse energy and pulse duration as well as the fiber length. We identified the regions of optimum parameters. We also proposed simple analytical considerations that help to identify the role of the Raman effect and obtain the approximate scaling of the optimal pulse duration. The numerical results were supported by experimental data.

\section{Polarization squeezing description and numerical modeling}
We focus on two-mode polarization squeezing because its experimental realization is quite robust and less susceptible to various technical disturbances. The scheme we consider both in our modeling and experiment utilizes propagation of two pulses with the orthogonal polarizations aligned along axes of a birefringent nonlinear fiber. Both pulses experience Kerr squeezing. The polarization squeezing relies on the fact that the quantum uncertainty of the polarization state of two properly combined Kerr squeezed states can in some direction be made smaller than the shot- noise limit. Polarization state and polarization fluctuations can be described in terms of the Stokes operators
\begin{align}
\hat{S}_0 &= \hat{a}_H^{\dagger} \hat{a}_H + \hat{a}_V^{\dagger} \hat{a}_V, \qquad 
\hat{S}_1 = \hat{a}_H^{\dagger} \hat{a}_H - \hat{a}_V^{\dagger} \hat{a}_V, \nonumber \\
& \\
\hat{S}_2 &= \hat{a}_H^{\dagger} \hat{a}_V + \hat{a}_V^{\dagger} \hat{a}_H, \qquad
\hat{S}_3 = i (\hat{a}_V^{\dagger} \hat{a}_H - \hat{a}_H^{\dagger} \hat{a}_V),\nonumber
\end{align}
where $\hat{a}_{H/V}^{\dagger}$ and $\hat{a}_{H/V}$ are creation and annihilation operators of two field modes, corresponding to orthogonal horizontal/vertical polarization modes.  

The uncertainty relations for the polarization operator and the corresponding squeezing can be defined in an SU(2)-invariant manner~\cite{Korolkova}. The operators $\hat{S}_{1,2,3}$ can be represented as Cartesian components of a Stokes operator vector $\hat{\boldsymbol{S}}=(\hat{S}_1,\hat{S}_2,\hat{S}_3)$, and $\hat{S}_{0}$ represents the total photon number. We can define squeezing without explicit use of Cartesian projections of the Stokes operator vector, by introducing the component $S_{\parallel}$  parallel to the mean value $\langle \hat{\boldsymbol{S}} \rangle$   and two components  $\hat{S}_{\perp 1}$, $\hat{S}_{\perp 2}$ in the plane orthogonal to  $\langle \hat{\boldsymbol{S}} \rangle$ (the so-called ``dark plane")~\cite{Corney2008}. The nontrivial uncertainty relation for variances   $\Delta^2 \hat{S}_{\perp 1}$, $\Delta^2 \hat{S}_{\perp 2}$ then reads as $\Delta^2 \hat{S}_{\perp 1} \Delta^2 \hat{S}_{\perp 2} \geq |\langle \hat{S}_{\parallel} \rangle|^2 $ . The squeezing is observed if there are components in the dark plane that obey~\cite{Korolkova}
\begin{equation}
\Delta^2 \hat{S}_{\perp 1} <|\langle \hat{S}_{\parallel} \rangle|<\Delta^2 \hat{S}_{\perp 2}. 
\end{equation}

The SU(2) invariance implies that the rotations of the polarization states, which can be done with the use of birefringent plates and polarization splitting and combining optics, do not destroy the polarization squeezing, provided losses are small. Then the squeezing can be measured after appropriate rotation of the polarization state and measurement of the Stokes parameter $\hat{S}_1$ by using a polarization splitter and a balanced detector~\cite{Corney2008}. Moreover, this quantum polarization description can be mapped one-to-one onto the quantum description of SU(2) interferometers~\cite{yurkePRA86}, for which it is known that the sensitivity can be enhanced by using squeezed light states~\cite{Caves81}. This means that polarization squeezed light can be used for precision interferometric measurements. It was shown that bright squeezed light can be used for increasing the precision of polarimetry~\cite{Rudnicki}, and enhancing the sensitivity of polarization interferometer~\cite{KalininICLO}. 

Efficient numerical modeling of quantum dynamics leading to squeezed-state formation in the fiber requires certain assumptions and simplifications. We assume that the pulses propagate independently of each other in two polarization modes of the fiber. We apply the truncated Wigner method to model the quantum dynamics. This method is based on reconstructing the Wigner distribution by gathering a large number of stochastic trajectories using the stochastic nonlinear Schrödinger equation~\cite{Drummond2001, Corney2001, Bonett2021, Sorokin2021}. Our particular implementation of this equation takes into account fiber dispersion (up to the third order) and the nonlinear response mediated by both the Raman and instantaneous electronic interactions. We model this equation with the parameters of a particular fiber which was used in our experiment (second-order dispersion $\beta_2=-10.5$ps$^2$/km, third- order dispersion $\beta_3=0.155$~ps$^3$/km, nonlinear coefficient $\gamma=3$~W$^{-1}$km$^{-1}$, and the Raman response function as in~\cite{Sorokin2021}). The pulse parameters were chosen in the ranges covering the values accessible in our experiment. We adopt the polarization squeezing detection scheme and used the corresponding routine to calculate the squeezing from the numerically simulated data~\cite{Corney2008, Sorokin2021}. 

In numerical modeling we calculate the squeezing for various input pulse parameters and various fiber lengths. We prepared initial conditions in the form of hyperbolic secant-shaped pulses $A=A_0/\cosh{(t/\tau)}$  with different durations in the range $T=0.11-0.5$~ps ($T$ is FWHM duration, $T=1.763\tau$) and with a pulse energy in the range $E=22.5-120$~pJ. We calculate the quantum dynamics for a propagation distance of up to 30 m for each initial condition. For each set of initial conditions, we modeled 5000 realizations of stochastic trajectories to reconstruct the squeezing ellipse. In the process of modeling, we calculated the squeezing at intermediate distances along the fiber and recorded the obtained values. After the calculation of squeezing, we could also introduce the losses of the detection scheme, which are inevitable in the experiment.

\section{Experimental study}
We carried out an experimental study of polarization squeezing, which allowed us to compare the measurements with our theory. The experimental setup for generating and measuring squeezing generally followed the schematic presented in~\cite{Corney2008}. The two-mode squeezed light needed to achieve polarization squeezing was generated in the polarization-maintaining fiber (3M~FSPM-7811). Femtosecond pulses at the central wavelength of 1.56 $\mu$m from a mode-locked laser with an adjustable pulse width and energy were launched into both polarization axes with equal power. The laser signal was shot-noise limited above radio frequencies of a few MHz. Because of the fiber birrefringence and the difference in the group velocities of the polarization modes, the pulses quickly separated in time and propagated in the fiber almost independently. To match the pulse arrival time at the output we used two consecutive fiber pieces of precisely equal length spliced together with swapped fast and slow axes (rotated by 90 degrees)~\cite{KalininICLO}. This allowed us to make the setup simple and robust eliminating the free-space interferometer required in the original scheme~\cite{Corney2008} to adjust the pulse arrival time. We tested two fiber lengths of 5.2 and 30 meters. To measure squeezing we first adjusted the polarization state and orientation of the squeezed ellipse using waveplates (as described in~\cite{Corney2008}) and measured the noise in the Stokes parameter $\hat{S}_1$ using a polarization beam splitter, a balanced photodetector and a radiofrequency spectrum analyzer. We used several laser settings providing different pulse durations and for each setting the pulse energy was optimized for the best squeezing.

\section{Results and analysis}
The analysis of numerical data allowed us to identify the most important processes and parameters affecting squeezing. The entire set of simulations provide a 3D data set, with squeezing calculated as a function of pulse duration, pulse energy, and fibre length.  The slices of the 3D data set showing the squeezing versus input pulse duration and energy at eight distances along the fiber are presented in Fig.~1. The simulation was carried out on a $14\times 14$ grid, but we have used data interpolation for a better visual representation. 

\begin{figure}[t]
\centering\includegraphics[width=\textwidth]{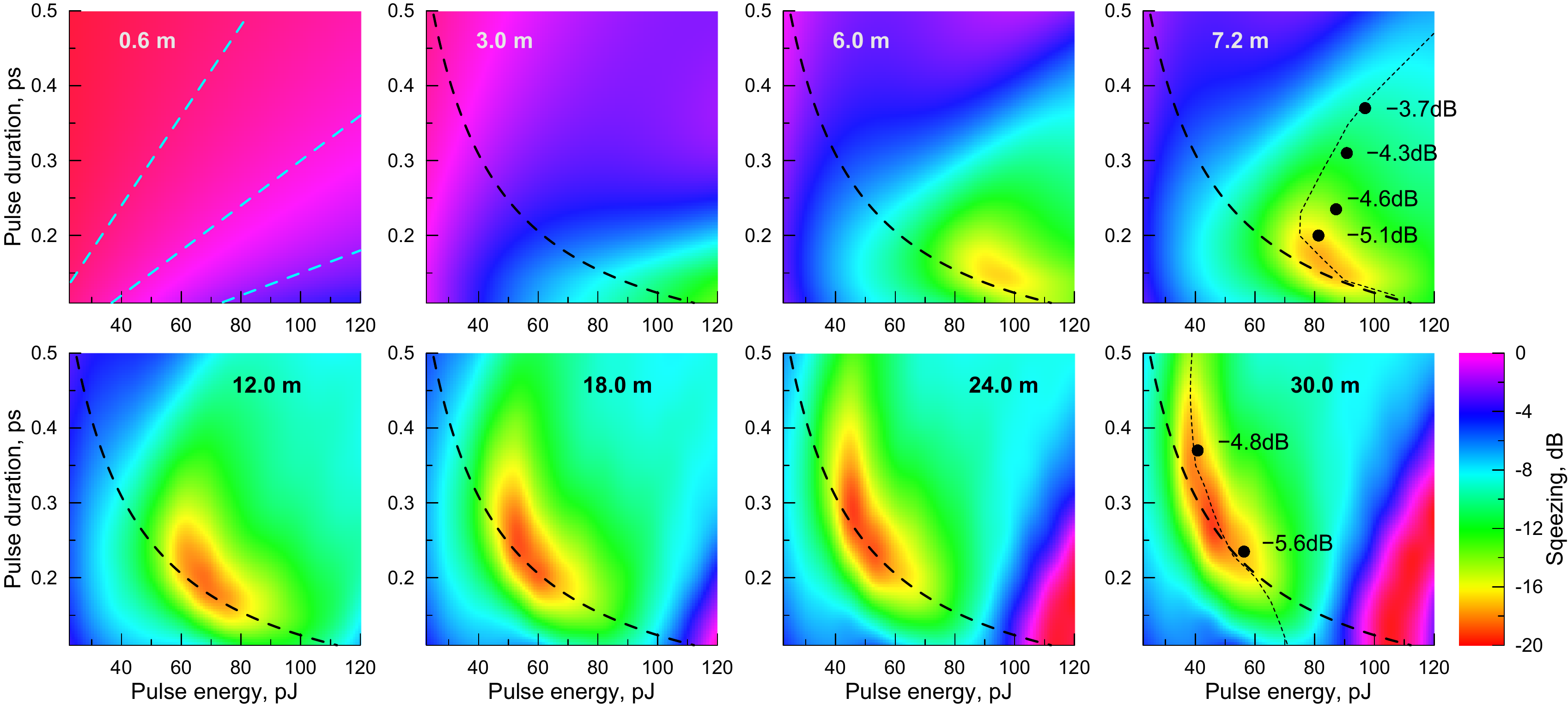}
\caption{Simulated squeezing for different pulse energies and durations at eight distances along the fiber from 0.6 to 30 meters (color maps). Dashed cyan lines in the plot for 0.6 m correspond to constant peak power. Dashed black lines in the rest of the plots represent fundamental soliton parameters. Dotted lines in plots for 7.2 m and 30 m correspond to the pulse energy maximizing squeezing for a given pulse duration. Black dots show data points obtained in the experimental optimization of the pulse energy. Measured squeezing values are shown next to each dot.}
\end{figure}

We can see that at the very beginning of the pulse propagation the squeezing mainly depends on the peak power of the pulse. This behavior is consistent with simple considerations: At small distances the pulse shaping effects are not pronounced, so the pulse mainly experiences self-phase modulation and hence acquires some squeezing proportional to the peak power and the fiber length. To emphasize this, we add lines of constant peak power to the plot. At larger distances the soliton effects begin to play an important role, so the pulse dynamics becomes more complicated. Pronounced regions of better squeezing are formed along curved lines. Better squeezing is observed for the pulse parameters close to the fundamental soliton. To demonstrate this, we plot dashed black lines, corresponding to the soliton parameters  $T=1.763\tau = 3.526|\beta_2|/\gamma E$. For two distances (7.2~m and 30~m) we also plot curves corresponding to the pulse energies maximizing squeezing at variable pulse duration (dotted lines in Fig. 1). Note that for small and intermediate fiber lengths and large durations, solitons do not have enough distance to form, so the squeezing for such pulses largely depends on the input pulse peak power. This results in a C-shaped optimal energy curve for the fiber length of 7.2~m.

We also compared our numerical findings with experimental results. The experimental points, obtained for several pulse durations after optimizing the pulse energy, are shown in Fig.~1 for the fiber lengths of 7.2~m and 30~m. It is evident that the experimental points align very well with the curves of optimum pulse energy obtained in numerical modeling. They represent two distinctive cases. For shorter distances, the optimum pulse energy increases as the pulse duration increases above $\sim$0.3 ps. For longer distances, the optimum pulse energy decreases as the pulse width increases so that the pulse parameters stay close to those of fundamental soliton. The absolute values of squeezing in the experiment are significantly smaller than the modeled ones, but this is because the simulation presented in Fig. 1 does not include losses. However, the trends in the optimal pulse parameters are similar, since the effect of losses on squeezing does not depend directly on the pulse energy or duration. 

We extracted the maximum squeezing and the corresponding pulse parameters for different fiber lengths, as shown in Fig.~2. It is seen that better squeezing can be achieved in longer fibers,  requiring progressively larger pulse durations and lower pulse energies. From the data we calculated the soliton number parameter $N^2= \tau \gamma E /2|\beta_2|$, which characterizes how close the pulse is to the fundamental soliton  ($N$=1 corresponds to the fundamental soliton). Note that at short propagation distances the best squeezing is observed at energies of about 10\% larger than the soliton energy. For large distances the best squeezing is achieved for pulses very close to the fundamental solitons.

Optical losses in the fiber, at the fiber output and in the squeezing detector can strongly limit the squeezing, especially for very large values demonstrated in lossless modeling. The effect of losses at the output and lower than unity efficiency of the detectors can be simply added on top of the quantum dynamics modeling~\cite{BachorBook}. The effect of distributed fiber losses needs to be directly modeled using the propagation equation, but it can also be included approximately as lumped losses at the output, and we did so to speed up our modeling. In Fig. 2 we show the squeezing calculated when different losses are included: intrinsic fiber losses of 1dB/km and losses at the fiber output and in the detection scheme. Including only the fiber losses, the squeezing is still very strong, although it starts to roll off with increasing fiber length. With other additional loss of 20\% (estimated value for our experiment) the observed squeezing saturates at around $-6$~dB, which is close to the experimentally measured result. With smaller additional losses of 5\%, which seems feasible in a carefully optimized experiment, we can expect observed squeezing at the level of $-10$ to $-12$ dB. 

\begin{figure}[t]
\centering\includegraphics[width=\textwidth]{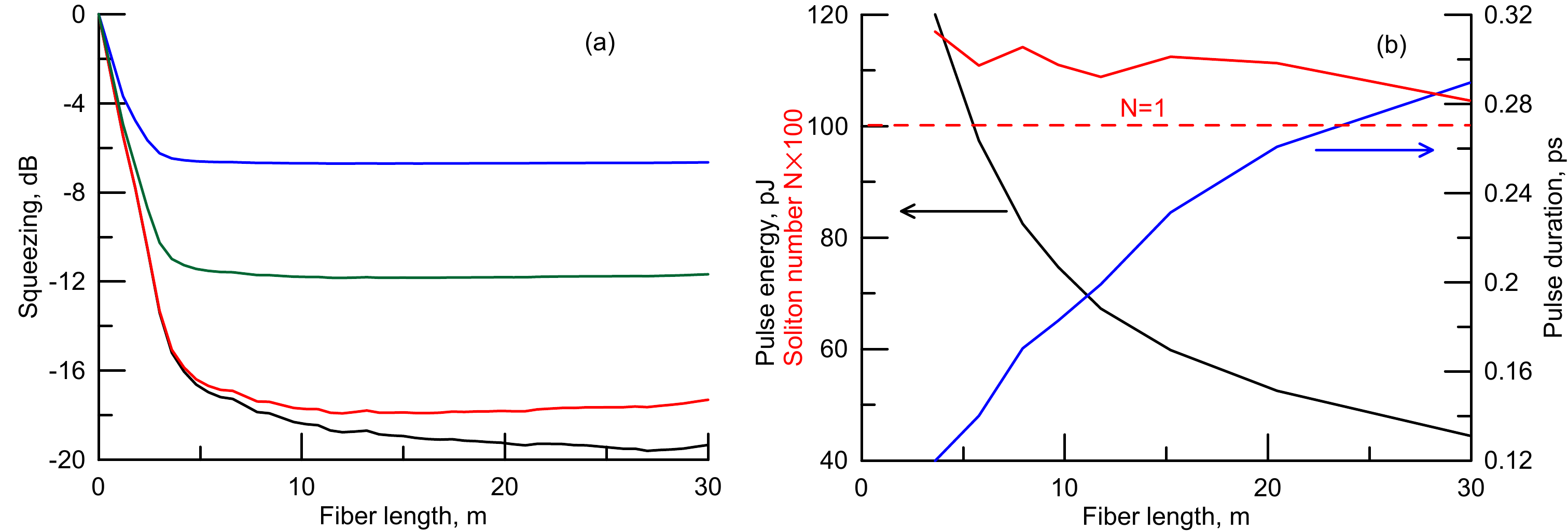}
\caption{Maximum squeezing as a function of the fiber length for different losses (a): no losses (black curve), intrinsic fiber losses only (red curve), fiber losses and external losses of 5\% (green curve), fiber losses and external losses of 20\% (blue curve). Optimal pulse parameters (b): energy (black curve, left axis), duration (blue curve, right axis), soliton number (red curve, left axis, multiplied by 100). }
\end{figure}

\section{Analysis of pulse duration limitations due to Raman effect}

Now we discuss the optimization of the pulse duration. From the simple picture of the squeezing building up due to the Kerr effect, one may expect that the squeezing would improve with increasing soliton energy (and corresponding shortening of its durations). At small propagation distances, regions of best squeezing are indeed observed for shorter durations and highest energies, but the optimum is gradually shifted towards longer durations and smaller energies at larger distances. For shorter pulses the squeezing degrades abruptly. We illustrate this in Fig. 3, in which we plot the maximum achievable squeezing optimized with respect to the pulse energy as a function of the pulse duration and the fiber length. A well-defined region of the best squeezing is observed. The optimum pulse duration shifts slowly towards larger values as the propagation distance increases.   

The observed behavior can be explained by evaluating the influence of nonconservative Raman effects. The Raman effect for ultrashort solitons manifests itself in gradual self-frequency shift of the pulse central frequency~\cite{Gordon}. The rate of soliton self-frequency shift is given by an approximate formula~\cite{Agrawal}
\begin{equation}
\frac{d\Omega}{d z} = \frac{8 T_R |\beta_2|}{15 \tau^4} \, ,
\end{equation}
where $T_R$ characterizes the strength of the Raman response, $T_R\sim 3-4$~fs depending on the particular shape of the response function. For the quantum evolution the influence of the Raman effect is more complicated, but some useful conclusions can be drawn based on the following considerations. The squeezing is related to certain correlations between the frequency side-bands of quantum noise. These correlations build up due to the Kerr effect during soliton propagation. The Raman effect redistributes the spectral components of the soliton and destroys these correlations. To be able to make an estimate, we assume that the correlations are destroyed and the squeezing is reduced when the soliton frequency spectrum is shifted by an amount comparable to the pulse spectral width. The FWHM spectral width $\Delta \Omega$ is inversely proportional to the pulse duration $\Delta \Omega \approx 2/T$. This leads to the condition
\begin{equation}
\frac{|\beta_2|T_R z }{T^3} \equiv K \ll 1,
\end{equation}
where the dimensionless coefficient $K$ absorbs all the constants. This condition must be well fulfilled so that the Raman effects can be neglected. Figur~3 shows lines of constant $K$. If we start from long pulse durations the squeezing starts to improve as the soliton duration decreases for any fixed fiber length corresponding to increasing $K$. However, as $K$ increases too much the squeezing saturates and then degrades. The contours of the best squeezing region coincide very well with the analytical predictions. The threshold at which the Raman effect becomes important corresponds to $K\sim 0.05 \ldots 0.1$. Although the presented numerical modeling was carried out for particular fiber parameters, the analytical condition (4) is fairly universal and thus could be used as a guide in planning and optimizing experiments. 

\section{Discussion and conclusion}

Our numerical modeling provides useful insights in how to optimize fiber polarization squeezing. The numerical results match the experimental observations fairly well in terms of the optimal combinations of pulse energy and pulse duration, for short as well as for long fiber length. However, the numerical results giving the best match were obtained for about fiber lengths differing by about 30\%. This can be explained by the fact that propagation in orthogonal polarization modes is not completely independent. Near the input and output ends of the fiber the pulses overlap in time, so the cross-Kerr interaction leads to an increase in the effective nonlinearity experienced by the pulses. The distance at which the pulses separate in time is about  0.5~m.  Along this distance, the cross-Kerr contribution induced by the orthogonal pulse with the same energy and peak power is added to the selfaction of the considered pulse~\cite{Agrawal}. In our simplified modeling we assumed independent propagation and neglected cross-Kerr effect, so longer fiber length was required to achieve similar effect. 

\begin{figure}[t]
\centering\includegraphics[width=7cm]{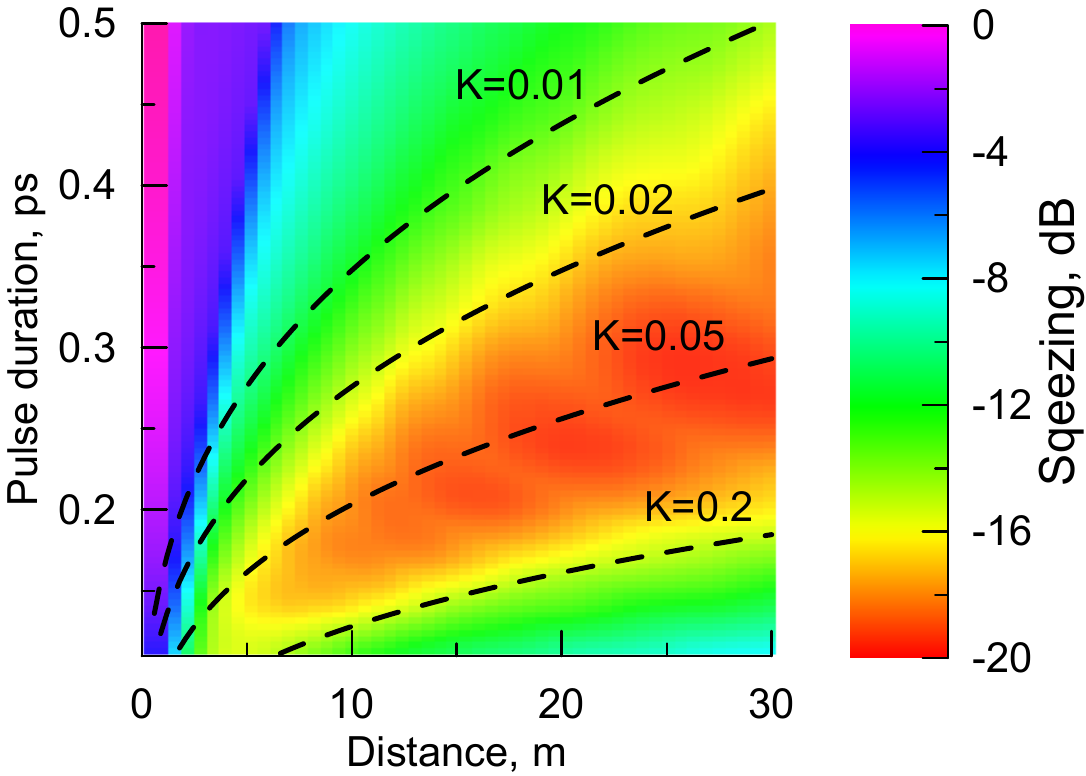}
\caption{Simulated squeezing optimized with respect to the pulse energy as a function of the pulse duration and the fiber length. Black dashed lines correspond to constant values of $K$ in (4).  }
\end{figure}

The experimentally measured squeezing is severely affected by losses. The squeezing saturates as it approaches the limit set by losses in the fiber and the detection scheme. However, the intrinsic fiber losses are quite low for considered fiber lengths and the theoretically achievable squeezing is quite strong even with these internal losses taken into account. So, the modeling presented here shows that a significant increase of the observed squeezing is realistic for a new experimental setup with largely reduced external losses. 

In conclusion, we performed the numerical simulation of polarization quantum squeezing in a nonlinear fiber aimed at the optimization of squeezing with respect to the pulse duration and energy as well as the fiber length and losses. Based on the analysis of the 3D data space obtained in the modeling we identified the parameter areas for the best squeezing and described general trends covering a wide range of pulse and fiber parameters. We proposed a simple analytical approximation, which takes into account the Raman effect and provides the optimal pulse duration for given fiber parameters.  

\begin{backmatter}

\bmsection{Funding}
Ministry of Science and Higher Education of the Russian Federation, contract  075-15-2022-316.

\bmsection{Disclosures} The authors declare no conflicts of interest.

\medskip

\bmsection{Data availability} Data underlying the results presented in this paper are not publicly available at this time but may be obtained from the authors upon reasonable request.

\bigskip

\end{backmatter}

%%%%%%%%%%%%%%%%%%%%%%% References %%%%%%%%%%%%%%%%%%%%%%%%%

%%%%%%%%%% If using BibTeX:
%\bibliography{citations}

\end{document}